\documentclass[aps,reprint,showpacs,superscriptaddress]{revtex4-1}
\usepackage{blindtext}
\usepackage{amssymb, amsmath,framed}

\usepackage{bm}
\usepackage[colorlinks=true,linkcolor=blue]{hyperref}
\expandafter\ifx\csname package@font\endcsname\relax\else
 \expandafter\expandafter
 \expandafter\usepackage
 \expandafter\expandafter
 \expandafter{\csname package@font\endcsname}
\fi
\def\bq{\begin{equation}}
\def\eq{\end{equation}}
\def\bqy{\begin{eqnarray}}
\def\eqy{\end{eqnarray}}

\def\calc{\mathcal{C}}

\begin{document}
\title{Multi-region relaxed Hall magnetohydrodynamics with flow}

\author{Manasvi Lingam}
\email{mlingam@princeton.edu}
\affiliation{Department of Astrophysical Sciences, Princeton University, Princeton, NJ 08544, USA}
\author{Hamdi M. Abdelhamid}
\email{hamdi@ppl.k.u-tokyo.ac.jp}
\affiliation{Graduate School of Frontier Sciences, The University of Tokyo, Kashiwanoha, Kashiwa, Chiba 277-8561, Japan}
\affiliation{Physics Department, Faculty of Science, Mansoura University, Mansoura 35516, Egypt}
\author{Stuart R. Hudson}
\email{shudson@pppl.gov}
\affiliation{Princeton Plasma Physics Laboratory, PO Box 451, Princeton, NJ 08543, USA}

\begin{abstract}
The recent formulations of multi-region relaxed magnetohydrodynamics (MRxMHD) have generalized the famous Woltjer-Taylor states by incorporating a collection of ``ideal barriers'' that prevent global relaxation, and flow. In this paper, we generalize MRxMHD with flow to include Hall effects (MRxHMHD), and thereby obtain the partially relaxed counterparts of the famous double Beltrami states as a special subset. The physical and mathematical consequences arising from the introduction of the Hall term are also presented. We demonstrate that our results (in the ideal MHD limit) constitute an important subset of ideal MHD equilibria, and we compare our approach against other variational principles proposed for deriving the partially relaxed states.
\end{abstract}

\maketitle

\section{Introduction} \label{SecIntro}
Amongst all fluid descriptions of plasmas, none has proven to be as simple, relevant and versatile as ideal magnetohydrodynamics (MHD). Consequently, it has been widely employed in modelling fusion \cite{Fre14} and astrophysical \cite{Kuls05} plasmas. One of the most crucial aspects of ideal MHD entails the study of its equilibria. A great deal of attention has been centered around the idea proposed by \citet{KK58}, in that the plasma energy can be extremized subject to certain constraints, namely the so-called ``ideal-constraints'' that prevent topological variations of the magnetic field. In contrast, the Woltjer-Taylor state \cite{W58,T74,Tay86} is obtained by extremizing the magnetic energy, subject to holding \emph{only} the global magnetic helicity fixed (and constraints on the global flux and a boundary condition). This allows a wider class of equilibrium solutions to be accessed. We note that generalizations of the Woltjer-Taylor state to include flow have also been widely studied; see e.g. \cite{Wo58,Sud79,FA83,HB87} for some early treatments of this subject. 

The crucial assumption most commonly invoked in computing 3D equilibria in ideal MHD (for e.g., the VMEC code \cite{HW83,HvRM86}) is the existence of continuously nested flux surfaces. It is possible to relax this assumption, insisting instead that only a finite number of flux surfaces are present, thereby constituting a case of \emph{partial} relaxation. A model that gives rise to such equilibria can be seen as the generalization of the Taylor model, and originated in the studies undertaken by Hole, Hudson \& Dewar \emph{et al}. \cite{HHD06,HoHD07,HHD07}, and was dubbed ``multi-region relaxed MHD'' (MRxMHD). Subsequently, MRxMHD has been studied extensively, with a view towards understanding and extending it, in the works of \cite{DHMMH08,HMHD09,MHD09,HDHM12,Det13,DHDH14,DHH14,DYBH15}. The stepped pressure equilibrium code (SPEC) \cite{Het12}, which is based on MRxMHD, has been subsequently employed in multiple contexts, such as reverse field pinches \cite{DHet13}, magnetic islands and current sheets \cite{LHBH15,Let15}, and pressure-driven amplification \cite{Let16}.

A common feature of ideal MHD and MRxMHD is that they rely on the same physical principles (choice of invariants, etc.) in arriving at the corresponding relaxed states. It is, however, important to note that fluid models more encompassing than ideal MHD are existent in the literature. The most widely studied amongst them is Hall MHD \cite{Light60,Tur86,MY98,MY00,Yosh02,Aul11,Ket12,KSM12,BSAML14}, but equivalent treatments of two-fluid \cite{SI97,SI98,Ham14} and multi-fluid \cite{OT95,ML15} models can also be found. Such fluid effects play an important role in certain regimes, especially in space and astrophysical plasmas \cite{Hub95}. It is, thus, natural to formulate relaxation theories for these models along the lines of Woltjer and Taylor. 

Given the existence of two complementary approaches that generalize the notion of Taylor relaxation in different ways, viz. MRxMHD and Hall MHD, it is natural to look for a relaxation theory that encompasses both approaches. This is, indeed, the primary objective of this paper -- to construct a MRxMHD theory with Hall effects, which we christen henceforth as multi-region, relaxed, Hall MHD, or MRxHMHD. Owing to the property that Hall MHD is a singular perturbation of ideal MHD, we show that the partially relaxed states obtained from MRxHMHD are quite different from their MHD counterparts derived in \citet{DHDH14,DHH14}. This aspect is discussed in Sec. \ref{SecRelS}, where a detailed comparison of the two approaches is presented.

The outline of the paper is as follows. The relevant background material for carrying out the variational principle is presented in Section \ref{SecMPRP}. A detailed variation is carried out in Section \ref{SecRelS}, leading to the final (partially) relaxed states of MRxHMHD, and we also present the physical and mathematical effects that arise on account of the Hall term. We compare the MRxMHD states with flow against ideal MHD equilibria, and offer a few general comments, in Section \ref{SecMRXComp}. Finally, we conclude with a summary of our results and prospects for future work in Section \ref{SecConc}.

\section{Mathematical preliminaries and the variational principle} \label{SecMPRP}
In this Section, we shall present some of the relevant mathematical properties of Hall MHD and then set up the procedure to obtain the relaxed states of MRxHMHD. 

\subsection{The equations and properties of Hall MHD} \label{SSecMath}
The governing equations of Hall MHD are
\begin{equation}
\label{Cont}
\frac{\partial \rho}{\partial t}=-\nabla\cdot\left(\rho\textbf{V}\right),
\end{equation}
\begin{equation}
\label{Mom}
\frac{\partial \textbf{V}}{\partial t} + {\bf V}\cdot\nabla {\bf V} = \rho^{-1} \nabla p + \rho^{-1} \left(\nabla\times\textbf{B}\right)\times\textbf{B},
\end{equation}
\begin{equation}
\label{Ohm}
\frac{\partial \textbf{B}}{\partial t}=\nabla\times\left(\textbf{V}\times\textbf{B}\right)- d_{i}\nabla\times\left(\rho^{-1} \left(\nabla\times\textbf{B}\right)\times\textbf{B}\right).
\end{equation}
The equations have been normalized in Alfv\'enic units, such that $d_i = \lambda_i/L$ is the normalized ion skin depth, where $\lambda_i = c/\omega_{pi}$ is the ion skin depth in fiducial units and $L$ is the appropriate scale length. The total pressure, $p$, is assumed to be barotropic and adiabatic, i.e. it obeys the relation $p = \sigma \rho^\gamma$, where $\gamma$ is the adiabatic index and $\sigma$ is a proportionality constant. The dynamical variables of interest are $\rho$, ${\bf V}$ and ${\bf B}$, being the total mass density, the center-of-mass fluid velocity and the magnetic field respectively. 

It is worth remarking that Hall MHD has a (noncanonical) Hamiltonian formulation \cite{YH13,AKY15}. The Hamiltonian formulation of Hall MHD is particularly useful in extracting a special class of invariants known as the Casimirs \cite{AKY15,LMM15}, which are
\begin{equation}
\label{Mass}
M=\int_{\Omega}\rho~d^{3}\tau,
\end{equation}
\begin{equation}
\label{MagHel}
K=\frac{1}{2}\int_{\Omega}\textbf{A}\cdot\textbf{B}~d^{3}\tau,
\end{equation}
\begin{eqnarray}
\label{CanHel}
C&=&\frac{1}{2}\int_\Omega \left(\textbf{A}+d_{i}\textbf{V}\right)\cdot \left(\textbf{B}+d_{i}\nabla\times\textbf{V}\right)~d^{3}\tau \nonumber \\
&\equiv& \frac{1}{2}\int_\Omega \textbf{P}\cdot\left(\nabla \times \textbf{P}\right)~d^{3}\tau,
\end{eqnarray}
where $\textbf{P}=\textbf{A}+d_{i}\textbf{V}$ and is, sometimes, referred to as the (ion) canonical momentum. Similarly, we note that $C$ is often referred to as the canonical helicity or the generalized helicity \cite{Tur86,MY98}. 

A few other points, which we shall call upon later, must also be stated:
\begin{enumerate}
    \item If we introduce the electron velocity ${\bf V}_e = {\bf V} - d_i \nabla \times {\bf B}/\rho$, it is easy to show that (\ref{Ohm}) becomes akin to the ideal MHD induction equation except for ${\bf V} \rightarrow {\bf V}_e$. As a result, it is still viable to speak of the magnetic flux being conserved, but it is advected along the electron `particle' trajectory (in the Lagrangian picture) \cite{KLMWW14,LMM16}.
    \item If we replace ${\bf V}$ with ${\bf V}_e$ in (\ref{Cont}), it is easy to verify that the expression remains unchanged. This follows from the vector identity that the divergence of a curl vanishes. As a result, one may imagine even the density being advected along the electron trajectory. 
    \item In \citet{Tur86} and \citet{MY98}, it was pointed out that Hall MHD (in the barotropic or incompressible limit) could be cast into a pair of coupled vorticity-type equations. Based on this result, it is possible to construct a canonical vorticity flux $\int \nabla \times {\bf P} \cdot {\bf n} \,d^2\sigma$, which is advected along the \emph{ion} trajectory in the Lagrangian formalism \cite{DAML16}.
    \item Thus, it is viable to consider all variables being advected along the electron trajectory, except for the composite variable ${\bf P}$ that is advected along the ion trajectory \cite{DAML16}.
\end{enumerate}
The key difference, from the perspective of Lagrangian action principles, between Hall MHD and ideal MHD stems from the differences in the advection of the magnetic field (see Sec. III of \cite{DAML16}), which necessitates the introduction of another Lagrangian variable (and trajectory). The presence of the ion skin depth can be interpreted as arising from this particular feature of the magnetic field in Hall MHD. In what follows, we describe an Eulerian variational principle, but we wish to emphasize that a Lagrangian approach along the lines of \citet{DYBH15} could also be adopted instead. Since the ion skin depth or the ``barrier'' width aren't explicitly present in the action, we anticipate that they would not critically impact the properties of MRxHMHD. 

However, a clearer picture of the effects engendered by the Hall term would likely emerge when a non-zero barrier width is incorporated into the model in a self-consistent manner. We shall not pursue this line of investigation here, but we intend to address this matter in future publications. The term `barriers', introduced earlier, refers to the so-called ``ideal'' transport barriers that separate the different plasma regions, and they are assumed to be the magnetic flux surfaces in MRxMHD \cite{DHet13}, and will be discussed in due course.

\subsection{The MRxHMHD variational principle}
We shall consider a plasma system that consists of $N$ nested plasma regions $\left(R_{l}\right)$ separated by the ideal barriers. The energy is 
\begin{equation}
\label{Energy}
E=\sum_{l}E_{l}=\int_{R_{l}}\left\{\rho \frac{V^{2}}{2}+\frac{\sigma_{l}}{ \gamma-1}\rho^{\gamma}+\frac{B^{2}}{2}\right\} d^{3}\tau,
\end{equation}
whilst the mass, magnetic helicity and canonical helicity carry over from (\ref{Mass}), (\ref{MagHel}) and (\ref{CanHel}) respectively. In the multi-region picture, they correspond to
\begin{equation}
\label{MassMR}
M_{l}=\int_{R_{l}}\rho~d^{3}\tau,
\end{equation}
\begin{equation}
\label{MagHelMR}
K_{l}=\frac{1}{2}\int_{R_{l}}\textbf{A}\cdot\textbf{B}~d^{3}\tau,
\end{equation}
\begin{equation}
\label{CanHelMR}
C_{l} = \frac{1}{2}\int_{R_{l}} \textbf{P}\cdot\left(\nabla \times \textbf{P}\right)~d^{3}\tau.
\end{equation}
In addition to the above constraints, we also consider the toroidal component of the angular momentum, following the approach of \citet{DHDH14,DHH14}, as an additional constraint. Thus, we have
\begin{equation}
\label{AngMom}
L_{l}=\widehat{\textbf{z}}\cdot\int_{R_{l}}\rho \textbf{r}\times\textbf{V}~d^{3}\tau=\int_{R_{l}}\rho R~\textbf{V}\cdot \widehat{\phi}~d^{3}\tau,
\end{equation}
where $R$ denotes the cylindrical radius. There are subtleties associated with toroidal angular momentum conservation, and we refer the reader to \citet{DHDH14,DHH14} for a detailed discussion of the same. A boundary (interface) that remains axisymmetric during the relaxation process will ensure the conservation of the toroidal angular momentum \cite{DHDH14}. As this is undoubtedly a strong constraint, one can drop it if necessary, but the basic thrust of the analysis is not affected. It is worth pointing out that 3D MHD equilibria such as ``snakes'' \cite{Del13} cannot be modeled if we operate under the assumption that $L_{l}$ is conserved.

We also need to specify the boundary conditions for our system. These conditions arise from the flux constraints, viz. the conservation of the magnetic flux and the canonical vorticity flux (see Point 3 of Sec. \ref{SSecMath}). The relevant details, in the case of ideal MHD, can be found in Sec. IV of \citet{SLK01}. When it comes to Hall MHD, the boundary conditions correspond to ${\bf B}\cdot{\bf n} = 0$ and $\nabla \times {\bf P} \cdot {\bf n} = 0$; subtracting the former from the latter leads to $\nabla \times {\bf V}\cdot{\bf n} = 0$ \cite{Ket12}. Following the approach outlined by \cite{SLK01}, the flux constraints of Hall MHD translate to the equivalent conditions
\begin{equation} \label{BCMag}
\left(\textbf{n}\times\delta\textbf{A}\right)=-\left(\textbf{n}\cdot\xi_{e}\right)\textbf{B},
\end{equation}
\begin{equation} \label{BCCan}
\left(\textbf{n}\times\delta\textbf{P}\right)=-\left(\textbf{n}\cdot\xi_{i}\right)\nabla \times \textbf{P}.
\end{equation}
Note that $\xi_i$ and $\xi_e$ stand for the ion and electron displacements respectively. We observe that (\ref{BCCan}) involves $\xi_i$ since the variable ${\bf P}$ exhibits ion advection, as noted in Section \ref{SSecMath}. Moreover, we also wish to point out the fact that (\ref{BCMag}) and (\ref{BCCan}) are \emph{not} arbitrary. They are consequences of the frozen-in flux constraints, except that the former and latter are advected along different fluid trajectories. 

The energy functional of the MRxHMHD reads as
\begin{eqnarray}
\label{EF}
W&=&\sum_{l}\Big\{E_{l}-\nu_{l} \left(M_{l}-M_{l}^{0}\right)-\frac{1}{2}\mu_{l} \left(K_{l}-K^{0}_{l}\right) \nonumber \\
&&\quad -\lambda_{l} \left(C_{l}-C^{0}_{l}\right)-\Omega_{l} \left(L_{l}-L^{0}_{l}\right)\Big\},
\end{eqnarray}
where $\nu_{l}$, $\mu_{l}$, $\lambda_{l}$ and $\Omega_{l}$ are the Lagrange multipliers used to enforce the constraints on the mass, magnetic helicity, canonical helicity, and the angular momentum in each sub-volume. Here, the $X^0_l$'s are the constrained values of the respective $X_l$'s.

The expression for the magnetic helicity, given by (\ref{MagHelMR}), is not gauge invariant. To ensure gauge invariance, one can include two loop integrals that encapsulate the amount of toroidal/poloidal flux contained within each region, and the loop integrals are computed about the inner (outer) boundary of a given region in the poloidal (toroidal) direction. For more details, the reader is referred to \cite{Det13,DHDH14}. However, it turns out that these new terms, even upon inclusion, do not contribute to the final result, and we have omitted them in our analysis for the sake of clarity. A similar line of reasoning is also valid when dealing with the canonical helicity $C_l$. In both these respects, we follow the approach employed by \cite{Yosh02} in their formulation and analysis of variational principles for two-fluid plasmas.

\section{Derivation of the partially relaxed states with Hall effects} \label{SecRelS}
In this Section, we shall use (\ref{EF}) as the functional subject to extremization, viz. we compute $\delta W = 0$. By doing so, we expect to compute the constrained minimum energy states, but it must be recognized that a rigorous analysis will also necessitate taking the second variation of $W$ to ensure that the final result is, indeed, a minima \cite{HMRW85,Mor98,Mor05}. Alternatively, one can also carry out a numerical analysis to determine the stability of these equilibria.

Our approach belongs to the category of variational principles that extremize the energy. An alternative approach, also widely studied in the literature, is to extremize the entropy instead \cite{HB87,Ham14} and a detailed discussion of this subject can be found in \citet{DHMMH08}.

Before proceeding further, we begin by noting a useful identity \cite{SLK01,Sp03,DHDH14} that we shall use in the subsequent derivations. It corresponds to
\begin{equation}
\delta \int_{D} X d^{3}\tau= \int_{D} \delta X d^{3}\tau+ \int_{\partial D} \left(\textbf{n}\cdot \xi \right)X d^{2}\sigma,
\end{equation}
where $D$ and $\partial D$ are the volume and bounding surface respectively, $X$ is the functional under consideration, and $\xi$ corresponds to the fluid displacement.

\subsection{The Energy functional}
The first variation of the energy functional $E_l$ can be obtained as follows
\begin{equation}
\delta E_{l}=\delta\int_{R_{l}}\rho \frac{V^{2}}{2}d^{3}\tau+\delta\int_{R_{l}}\frac{\sigma_{l}}{ \gamma-1}\rho^{\gamma}d^{3}\tau+\delta\int_{R_{l}}\frac{B^{2}}{2} d^{3}\tau,
\end{equation}
and the individual components yield
\begin{eqnarray} \label{KEVar}
\delta\int_{R_{l}}\rho \frac{V^{2}}{2}d^{3}\tau&=&\int_{R_{l}}\delta\rho\frac{V^{2}}{2}d^{3}\tau+\int_{ R_{l}}\delta\textbf{V}\cdot\rho\textbf{V}~d^{3}\tau \nonumber \\
&&\,+ \int_{\partial R_{l}} \left(\textbf{n}\cdot \xi_{e} \right)\rho \frac{V^{2}}{2} d^{2}\sigma,
\end{eqnarray}

\begin{eqnarray} \label{PressVar}
\delta\int_{R_{l}}\frac{\sigma_{l}}{ \gamma-1}\rho^{\gamma}d^{3}\tau&=&\int_{R_{l}}\delta\rho\frac{\gamma}{ \gamma-1}\sigma_{l}\rho^{\gamma-1}~d^{3}\tau \nonumber \\
&&\, +\int_{\partial R_{l}}\left(\textbf{n}\cdot\xi_{e} \right)~\frac{\sigma_{l}}{ \gamma-1}\rho^{\gamma}~d^{2}\sigma,
\end{eqnarray}

\begin{equation} \label{MagVar}
\delta\int_{R_{l}}\frac{B^{2}}{2} d^{3}\tau=\int_{R_{l}}\delta\left(\frac{B^{2}}{2}\right) d^{3}\tau+\int_{\partial R_{l}}\left(\textbf{n}\cdot\xi_{e} \right)\frac{B^{2}}{2}~d^{2}\sigma,
\end{equation}
The first term in (\ref{MagVar}) can be further simplified as follows
\begin{equation} \label{MagVarT1}
\int_{R_{l}}\delta\left(\frac{B^{2}}{2}\right) d^{3}\tau=\int_{R_{l}}\left(\nabla\times\delta\textbf{A}\right)\cdot\left(\nabla\times\textbf{A}\right) d^{3}\tau,
\end{equation}
and we use the vector calculus identities
\begin{equation} \label{VCId1}
\nabla\cdot\left(\textbf{F}\times\textbf{G}\right)=\textbf{G}\cdot\left(\nabla\times\textbf{F}\right)-\textbf{F}\cdot \left(\nabla\times\textbf{G}\right),
\end{equation}
\begin{equation} \label{VCId2}
\int_{v}\nabla\cdot\textbf{F}~d^{3}\tau=\int_{\partial v}\textbf{n}\cdot \textbf{F}~d^{2}\sigma,
\end{equation}
in (\ref{MagVarT1}) to obtain the final relation
\begin{eqnarray} \label{MagVarT1Fin}
\int_{R_{l}}\delta\left(\frac{B^{2}}{2}\right) d^{3}\tau&=&\int_{R_{l}}\delta\textbf{A}\cdot\left(\nabla\times\textbf{B}\right) d^{3}\tau \nonumber \\
&&\, +\int_{\partial R_{l}}\left(\textbf{n}\times\delta\textbf{A}\right)\cdot\textbf{B}~d^{2}\sigma.
\end{eqnarray}
We are now free to substitute (\ref{BCMag}) into the second term of (\ref{MagVarT1Fin}). Next, we take the ensuing result and substitute it into (\ref{MagVar}). We obtain
\begin{equation} \label{MagVarFin}
\delta\int_{R_{l}}\frac{B^{2}}{2} d^{3}\tau=\int_{R_{l}}\delta\textbf{A}\cdot\left(\nabla\times\textbf{B}\right) d^{3}\tau - \int_{\partial R_{l}}\left(\textbf{n}\cdot\xi_{e} \right)\frac{B^{2}}{2}~d^{2}\sigma,
\end{equation}
and it is important to recognize that the second term on the RHS of (\ref{MagVar}) and (\ref{MagVarFin}) are the same, but \emph{opposite} in sign.

\subsection{The Helicity Functionals}
Let us first begin with the magnetic helicity. 
\begin{equation}
\delta K_{l}=\frac{1}{2}\int_{R_{l}}\delta\left(\textbf{A}\cdot\textbf{B}\right)~d^{3}\tau+\frac{1}{2}\int_{\partial R_{l}}\left(\textbf{n}\cdot\xi_{e}\right)\textbf{B}\cdot\textbf{A}~d^{2}\sigma,
\end{equation}
and we can simplify the first term on the RHS by invoking ${\bf B} = \nabla \times {\bf A}$ along with (\ref{VCId1}) and (\ref{VCId2}) and carrying out a procedure akin to that undertaken for the magnetic energy. Upon simplification, we end up with
\begin{equation}
\delta K_{l}=\int_{R_{l}}\delta\textbf{A}\cdot\textbf{B}~d^{3}\tau +\frac{1}{2}\int_{\partial R_{l}}{\bf A}\cdot\left[\textbf{n}\times\delta\textbf{A}+\left(\textbf{n}\cdot\xi_{e}\right)\textbf{B}\right]~d^{2}\sigma,
\end{equation}
and the second term on the LHS vanishes upon using the boundary condition (\ref{BCMag}). Thus, we have
\begin{equation} \label{MagHelFinal}
\delta K_{l}=\int_{R_{l}}\delta\textbf{A}\cdot\textbf{B}~d^{3}\tau.
\end{equation}
Dealing with the canonical helicity is much harder owing to its greater complexity. However, we can bypass a great deal of this complexity if we use the canonical momentum ${\bf P}$ as our composite variable. Furthermore, note that (\ref{BCCan}) dictates that the boundary conditions are most naturally interpreted in terms of ${\bf P}$ as well, since this condition arises from the conservation of the canonical vorticity flux. From (\ref{SSecMath}), we recall that ${\bf P}$ is advected along the ion trajectory, and that it is written in terms of ${\bf P}$ as per the second line of (\ref{CanHelMR}). Using these facts, we find that 
\begin{eqnarray} \label{CanIntVar}
\delta C_{l}&=&\frac{1}{2}\int_{R_{l}}\delta\left(\textbf{P}\cdot\left(\nabla \times \textbf{P}\right)\right)~d^{3}\tau \nonumber \\
&&\,+\frac{1}{2}\int_{\partial R_{l}}\left(\textbf{n}\cdot\xi_{i}\right)\textbf{P}\cdot\left(\nabla \times \textbf{P}\right)~d^{2}\sigma.
\end{eqnarray}
The second term on the RHS exhibits the label `$i$' since it's advected along the ion trajectory. Although ${\bf P}$ is comprised of two dynamical variables, we note that the two vector calculus identities (\ref{VCId1}) and (\ref{VCId2}) are still valid. Hence, we apply them to the first term of (\ref{CanIntVar}) and end up with
\begin{eqnarray}
\delta C_{l}&=&\int_{R_{l}}\delta\textbf{P}\cdot\left(\nabla \times\textbf{P}\right)~d^{3}\tau \nonumber \\
&&\, +\frac{1}{2}\int_{\partial R_{l}}{\bf P}\cdot\left[\textbf{n}\times\delta\textbf{P}+\left(\textbf{n}\cdot\xi_{i}\right)\nabla \times \textbf{P}\right]~d^{2}\sigma,
\end{eqnarray}
and by using (\ref{BCCan}) in the second term on the RHS of the above expression, we see that it vanishes identically. Hence, we end up with 
\begin{eqnarray}  \label{CanHelFinal}
\delta C_{l}&=&\int_{R_{l}}\Big[\delta\textbf{A}\cdot\left(\textbf{B}+d_{i}\nabla\times\textbf{V}\right) \nonumber \\
&&\quad \, +d_{i}\,\delta\textbf{V}\cdot\left(\textbf{B}+d_{i}\nabla\times\textbf{V}\right)\Big]~d^{3}\tau,
\end{eqnarray}
where we have used the fact that ${\bf P} = {\bf A} + d_i {\bf V}$ to rewrite our answer in terms of the dynamical variables. 

A comment regarding the variables ${\bf P}$ and ${\bf A}$ is in order at this juncture. The ion and electron trajectories are viewed as being distinct in the Lagrangian picture, and the advection of ${\bf P}$ and ${\bf A}$ occurs along the respective trajectories. However, the demands of locality, when operating in the Eulerian picture, complicate matters, as the two trajectories must be evaluated at the same physical `position'. A detailed discussion and explanation of this important issue can be found in \cite{KLMWW14} (see also \cite{DAML16}). We shall defer a more thorough investigation for subsequent publications, but we point out that an exact cancellation of the surface terms occurs for the two helicities, thereby leading us to (\ref{MagHelFinal}) and (\ref{CanHelFinal}). Owing to this outcome, our final results of our analysis are likely to remain unchanged for the most part.

\subsection{The mass and angular momentum functionals}
The variation of the angular momentum yields
\begin{eqnarray} \label{AngMomFinal}
\delta L_{l}&=&\int_{R_{l}}\delta\rho R~\textbf{V}\cdot \widehat{\phi}~d^{3}\tau+\int_{R_{l}}\delta\textbf{V}\cdot \rho R\widehat{\phi}~d^{3}\tau \nonumber \\
&&\, +\int_{\partial R_{l}}\left(\textbf{n}\cdot\xi_{e}\right)\left(\rho R~\textbf{V}\cdot \widehat{\phi}\right)~d^{2}\sigma.
\end{eqnarray}
The variation of the mass functional leads us to the result
\begin{equation} \label{MassFinal}
\delta M_{l}=\int_{R_{l}}\delta\rho~d^{3}\tau+\int_{\partial R_{l}}\left(\textbf{n}\cdot\xi_{e}\right)\rho~d^{2}\sigma,
\end{equation}

\subsection{The derivation of the relaxed states and jump condition} \label{SSecDerRS}
Our final expressions are given by (\ref{KEVar}), (\ref{PressVar}), (\ref{MagVarFin}), (\ref{MagHelFinal}), (\ref{CanHelFinal}), (\ref{AngMomFinal}) and (\ref{MassFinal}). Upon setting $\delta W = 0$, we obtain
\begin{equation}
\label{bVari}
\nabla\times\textbf{B}=\left(\mu_{l}+\lambda_{l}\right)\textbf{B}+ d_{i} \lambda_{l} \nabla\times\textbf{V},
\end{equation}
\begin{equation}
\label{VelVari}
\rho\textbf{V}= d_{i} \lambda_{l}\textbf{B}+d^{2}_{i} \lambda_{l}\nabla\times\textbf{V}+\rho \Omega_{l} R \widehat{\phi},
\end{equation}
\begin{equation}
\label{rhoVari}
\nu_{l}= \frac{1}{2}{V}^{2}+\frac{\gamma}{ \gamma-1}\sigma_{l}\rho^{\gamma-1}-\Omega_{l} R \textbf{V}\cdot\widehat{\phi},
\end{equation}
which have arisen from the variations with respect to ${\bf A}$, ${\bf V}$ and $\rho$ respectively. In addition, we also end up with a bevy of surface integral terms collectively given by
\begin{eqnarray}
&& \int_{R_{l}}\left(\textbf{n}\cdot \xi_{e} \right)\bigg\{\rho \frac{V^{2}}{2}+\frac{\sigma_{l}}{ \gamma-1}\rho^{\gamma}-\frac{{B}^{2}}{2} \nonumber \\
&&\hspace{0.8 in} -\nu_{l}\rho-\Omega_{l}\left(\rho R~\textbf{V}\cdot \widehat{\phi}\right)\bigg\}~d^{2}\sigma=0,
\end{eqnarray}
and we are free to substitute (\ref{rhoVari}) into the above expression. A lot of terms cancel leaving us with
\begin{equation}
\int_{R_{l}}\left(\textbf{n}\cdot \xi_{e} \right)\bigg\{-\sigma_{l}\rho^{\gamma}-\frac{{B}^{2}}{2}\bigg\}~d^{2}\sigma=0,
\end{equation}
and using the fact that $p_l = \sigma_l \rho^\gamma$, we arrive at the interface condition
\begin{equation}
\label{interface}
\left[\left[p_{l}+\frac{{B}^{2}}{2}\right]\right]= 0.
\end{equation}
At this stage, it is worth comparing our results with the ideal MHD counterparts obtained in \citet{DHDH14,DHH14}. The chief difference is that the canonical helicity of Hall MHD must be replaced by the cross helicity when dealing with ideal MHD. The rest of the invariants stay the same, implying that any differences that arise must be because of the difference between the cross and canonical helicity - we defer a detailed discussion of this topic to Sec. \ref{SSecHallRS}.

The partially relaxed states obtained above are similar to their (ideal) MHD counterparts in some ways.  For instance, it is easy to verify that the interface condition (\ref{interface}) is the same in both instances. This is not surprising since the interface condition is a manifestation of the force balance \cite{God04}, and it is well known that the momentum equation of ideal and Hall MHD are identical to one another \cite{Light60}. We also find that (\ref{rhoVari}) is exactly identical to the ideal MHD result. This is not surprising since the helicities do not contribute to this equation, and therefore we do not expect any differences in the final result. 

Next, let us consider (\ref{bVari}) and (\ref{VelVari}) and allow $d_i \rightarrow 0$. Upon comparing with \cite{DHDH14,DHH14}, we lose too many terms and the results do not match. At first glimpse, it may appear as though there was an error committed, but it is important to appreciate the mathematical fact that Hall MHD is a \emph{singular perturbation} of ideal MHD \cite{YMO04} and the cross helicity does \emph{not} follow from the canonical helicity simply by letting $d_i \rightarrow 0$, as indicated in \cite{OZ05,AKY15}.

Instead, let us take a closer look at (\ref{bVari}) and (\ref{VelVari}). If we introduce the new variable $\Lambda_l := d_i \lambda_l$ in these two equations, we find that they are identical to the ones presented in \cite{DHDH14,DHH14}, apart from the second term on the RHS of (\ref{VelVari}) which has the coefficient $d_i \Lambda_l$. The same result can be obtained without the need for this `reparametrization'. It is straightforward to verify that any linear combination of two Casimir invariants is also a Casimir, implying that
\begin{equation} \label{HCrossHel}
    \calc = \int_\Omega \left[\textbf{V}\cdot \textbf{B} + \frac{d_i}{2} \textbf{V} \cdot \left(\nabla \times \textbf{V}\right)\right]\,d^3\tau,
\end{equation}
is a Casimir invariant of Hall MHD, as it can be linearly constructed from (\ref{MagHel}) and (\ref{CanHel}). If we use (\ref{HCrossHel}), instead of (\ref{CanHel}), in the variational principle (\ref{EF}), the limit $d_i \rightarrow 0$ will duly yield the MRxMHD results delineated in \cite{DHDH14,DHH14}. We have chosen to use the canonical helicity in our variational principle, as it has been recently shown, by means of numerical simulations \cite{SS16}, to be a fairly robust invariant. In addition, the variable $\nabla \times {\bf P}$ that enters the canonical helicity is endowed with some interesting properties akin to the magnetic field (see Sec. \ref{SecMPRP} for further details), consequently simplifying the analysis.  

Lastly, we point out an interesting subset of our primary results. If we had started with an incompressible model, we would not have recovered (\ref{rhoVari}). Moreover, if our system did not conserve toroidal angular momentum (which can arise in certain circumstances, as discussed in \cite{DHDH14}), it amounts to setting $\Omega_l \rightarrow 0$ in (\ref{VelVari}) and (\ref{rhoVari}). Under these conditions, it is easy to verify that the resulting set of equations are the multi-region equivalent of the famous double Beltrami states obtained for incompressible Hall MHD in \cite{Tur86,MY98}. 

These multi-region counterparts, when simplified through the assumption of axisymmetry, yield a somewhat intricate set of equations for the field and flow profiles. The central equation corresponds to a generalized Grad-Shafranov equation endowed with Hall effects, thereby enabling some comparisons with other such studies \cite{MT01,Go04}. When we further restrict ourselves to 1D solutions (i.e. with only radial dependence), we find that the final solution is expressible as a linear combination of two independent solutions to the usual Grad-Shafranov equation. Thus, the analytic calculation, even in a simplified geometry (axisymmetry), reduces to a 2D (also higher-order) differential equation, which is significantly more complicated than the 1D equation which was solved in \cite{DHDH14,DHH14}. In view of the above reasons, we will not report on these results further, as we intend to cover these aspects in greater detail elsewhere. 

\subsection{The physical consequences of the Hall term} \label{SSecHallRS}
If the Hall effects are neglected, and we consider the ideal MHD limit, the system relaxes to a Woltjer-Taylor (single Beltrami) state along with the condition ${\bf V} \parallel {\bf B}$. The latter is often referred to as a field-aligned flow, since the flow is parallel to that of the magnetic field.  This feature has been a staple of ideal MHD based relaxation theories since the pioneering works by Woltjer in the 1950s \cite{W58,Wo58,Wo59}. 

On the other hand, replacing the cross helicity by the canonical helicity results in a non-zero component of the flow that is perpendicular to the magnetic field. Hence, we wish to emphasize that our model (or its variant) is particularly suited for modeling systems where this feature has been observed. We wish to point out that there exists sufficient numerical \cite{NYH04,BCE16,SS16} and experimental \cite{Ket09,Sar13,AKVJ13} evidence for flows that are not aligned with the fields, thereby implying that our model is likely to be of practical relevance in these contexts. Most of the aforementioned studies have also pointed out the role and importance of the Hall term (and the associated canonical helicity), as well as other non-ideal two-fluid effects, in regulating the emergent relaxed states. 

There are other effects, in addition to the absence of field-aligned flows, that result from the inclusion of Hall effects in relaxation theories. We have briefly remarked in Sec. \ref{SSecDerRS} that the double Beltrami states are a special class of solutions. A special feature of the double Beltrami states is that they can be written as a superposition of two single Beltrami fields. At certain critical points, it can be shown that the double Beltrami solutions collapse to a single Beltrami state. In this instance, the magnetic energy drops to a minimum, and the excess energy is transferred to the kinetic energy of the flow. For further details, both analytical and numerical, pertaining to this phenomenon, we refer the reader to \citet{MMNS01} and \citet{OSYM01}. 

The importance of this mechanism stems from its `catastrophic' nature \cite{MMNS01,OSYM01,OSYM02}, which makes it well suited for explaining explosive events; in this sense, it can be said to resemble reconnection, although the processes are quite different. We point out that this physical property, a staple of the double (and higher order) Beltrami states has been employed successfully in understanding coronal heating \cite{MMNS01}, the generation of stellar winds \cite{MNSY02,LingMa15}, and the formation of solar flares \cite{OSYM01,OSYM02,KM10}. We also refer the reader to associated studies of the H-mode boundary layer and the RT-1 experiment \cite{MY00,YMOIS01,GMY05,Yo10} which rely upon the special properties of the double Beltrami states, and they can be regarded, in a certain sense, as the relaxed states of Hall MHD. 

Amongst the other differences (in the physical aspects) brought about by the Hall term, we observe that the double Beltrami states play a role in the construction of the nonlinear Alfv\'en waves of Hall MHD \cite{KM04,Yos12,AY16,LinB16}. These waves are very different, both in form and properties, when compared against their ideal MHD counterparts, as they involve a ``nonlinearity-dispersion interplay'' \cite{AY16}. Thus, taken collectively, we believe that these examples constitute ample evidence of the important physical and mathematical differences brought about by the inclusion of the Hall term in self-organization and relaxed states.

\section{A note on the MR\lowercase{x}MHD equilibria} \label{SecMRXComp}
At this stage, we shall take a brief detour, and consider the MRxMHD equilibria derived by \citet{DHDH14,DHH14}. As described above, a \emph{careful} treatment of the partially relaxed states of MRxHMHD under the limit $d_i \rightarrow 0$ leads to the MRxMHD equilibria derived in \cite{DHDH14}. We choose to focus on MRxMHD (instead of MRxHMHD) as we are interested in understanding how the equilibria of \cite{DHDH14}, which give rise to partially relaxed states with flow, compare against ideal MHD equilibria. We shall also contrast these states against alternative approaches presented in the literature. 

\subsection{Ideal MHD equilibria with flow} \label{SSecMHDEqu}
Let us begin by writing down the expressions for ideal MHD equilibria endowed with flow.
\begin{equation} \label{ContEq}
    \nabla \cdot \left(\rho {\bf V}\right) = 0,
\end{equation}
\begin{equation} \label{MomEq}
\rho {\bf V}\cdot \nabla {\bf V} = {\bf J} \times {\bf B} - \nabla p,
\end{equation}
\begin{equation} \label{IndEq}
\nabla \times \left({\bf V} \times {\bf B}\right) = 0,
\end{equation}
where $p = \sigma \rho^\gamma$. It is easy to show that (\ref{MomEq}) can be rewritten as follows:
\begin{equation} \label{VelEq}
\boldsymbol{\omega} \times {\bf V} = \frac{{\bf J} \times {\bf B}}{\rho} - \nabla \left(\frac{\sigma \gamma\, \rho^{\gamma-1}}{\gamma - 1} + \frac{V^2}{2}\right),
\end{equation}
where we have introduced the notation $\boldsymbol{\omega} = \nabla \times {\bf V}$. 

\subsection{Partially relaxed states with flow} \label{SSecMRXFlow}
Here, we list the partially relaxed states with flow that were obtained in \cite{DHDH14}. As mentioned earlier, we can recover these states by taking the limit $d_i \rightarrow 0$ in our model, although there are some subtleties involved. The relevant equations are
\begin{equation} \label{TaylorRelax}
\nabla \times {\bf B} = \mu_l {\bf B} + \lambda_l \boldsymbol{\omega},
\end{equation}
\begin{equation} \label{AlfRelax}
\rho {\bf V} = \lambda_l {\bf B},
\end{equation}
\begin{equation} \label{BernRelax}
\sigma_l \frac{\gamma \rho^{\gamma-1}}{\gamma - 1} + \frac{V^2}{2} = \nu_l,
\end{equation}
and we note that the label `$l$' is present in the above equations, as we are looking at MRxMHD. In the continuum limit, this label can be dropped, and we shall do so henceforth for the sake of simplicity.

\subsection{Comparison of the two sets of equilibria}
We shall compare the results of Sec. \ref{SSecMRXFlow} against those of Sec. \ref{SSecMHDEqu}.

We begin by observing that (\ref{AlfRelax}) can be expressed as ${\bf V} \parallel {\bf B}$, or ${\bf V} \times {\bf B} = 0$. When this condition is satisfied, it is easy to verify that (\ref{IndEq}) is automatically satisfied. Similarly, if we take the divergence of (\ref{AlfRelax}), we end up with $\nabla \cdot \left(\rho {\bf V}\right) = 0$ on account of $\nabla \cdot {\bf B} = 0$. This condition is exactly identical to (\ref{ContEq}). We turn our attention to (\ref{VelEq}) now, and (\ref{BernRelax}) ensures that the second term on the RHS of (\ref{VelEq}) vanishes, i.e. the term inside the brackets. The remainder of (\ref{VelEq}) is given by 
\begin{equation} \label{VelEqTrunc}
\boldsymbol{\omega} \times {\bf V} = \frac{{\bf J} \times {\bf B}}{\rho},
\end{equation}
and we shall show that (\ref{TaylorRelax}) and (\ref{AlfRelax}) lead to the above relation. Let us take the cross product of ${\bf V}$ with (\ref{TaylorRelax}). This leads us to
\begin{equation} \label{JBOmV}
{\bf J} \times {\bf V} = \mu {\bf B} \times {\bf V} + \lambda \boldsymbol{\omega} \times {\bf V},
\end{equation}
and we invoke the expression for ${\bf V}$, in terms of ${\bf B}$, which is given by (\ref{AlfRelax}). We substitute this expression into the LHS and the first term on the RHS of (\ref{JBOmV}). This leads us to
\begin{equation}
\lambda \frac{{\bf J} \times {\bf B}}{\rho} = \lambda \boldsymbol{\omega} \times {\bf V},
\end{equation}
which is clearly identical to (\ref{VelEqTrunc}). 

Thus, the purpose of this exercise is now complete. We have shown that the equilibria derived by \cite{DHDH14} form a \emph{valid subset} of ideal MHD equilibria. For this reason, it is plausible that the partially relaxed states derived in \cite{DHDH14} (as well as the generalized states presented herein) constitute a physically meaningful set of MRxMHD equilibria with flow.

A few general observations regarding these partially relaxed states are in order. By substituting (\ref{AlfRelax}) into (\ref{TaylorRelax}), we find that
\begin{equation} \label{DefTaylor}
\nabla \times {\bf B} = \mu {\bf B} + \lambda^2 \nabla \times \left(\frac{{\bf B}}{\rho}\right), 
\end{equation}
which is clearly a deformation of the Taylor state since ${\bf J} \times {\bf B} \neq 0$. In fact, we find that a near-Taylor state is recovered only in two limits that are outlined below.
\begin{itemize}
\item When $|{\bf V}| \ll |{\bf B}|$, we can drop the last term on the RHS of (\ref{TaylorRelax}). This leads to a Taylor state to leading order.
\item When the system is nearly incompressible, this ensures that $\rho \rightarrow \mathrm{const}$ in (\ref{DefTaylor}), which in turn leads to ${\bf J} \times {\bf B} \rightarrow 0$.
\end{itemize}

The variational principles constructed herein, and in \cite{DHDH14}, were Eulerian in nature. A different variational formulation was presented in \cite{DYBH15} that relied upon the use of Lagrangian variables and induced variations. Although the same interface condition, namely (\ref{interface}), was recovered, there were some differences in the two approaches. The final expressions in \cite{DYBH15} corresponded to the Taylor state and the Euler equation for an ideal (neutral) fluid. It is straightforward to show that these equations also represent a valid set of the ideal MHD equilibria discussed in Sec. \ref{SSecMHDEqu}. However, the relations obtained in \cite{DYBH15} do not match the ones derived in \cite{DHDH14}, since the latter does not lead to a Taylor state, except under certain conditions.

The differences probably stem from the fact that ${\rho}$, ${\bf V}$ and ${\bf B}$ are treated as independent variables in the Eulerian picture. This is in sharp contrast to the Lagrangian treatment presented in \cite{DYBH15}, where the variations in ${\rho}$ and ${\bf V}$ are expressed in terms of the displacement as per the methodology described in \cite{FR60} (see also \cite{LM14}). On the other hand, ${\bf B}$ and $p$ are independent, and their variations are considered separately; see Eq. (3.21) of \cite{DYBH15} for a discussion of the same. 

The presence of induced variations also eliminated the need for the cross helicity (or, in our case, the canonical helicity) to be included in the variational principle. We note that this is quite different from most standard treatments in the literature, see e.g. \cite{Wo58,FA83,SI97,Yosh02,Ham14}. It is likely that a clearer picture will emerge once the SPEC code \cite{Het12} has been modified to implement flow. It will then be possible to compare the two approaches against experiments, or simulations from other sources, and thereby deduce their relative merits.

\section{Discussion and Conclusion} \label{SecConc}
The Woltjer-Taylor states of ideal MHD have proven to be widely successful in a host of fusion, space and astrophysical plasma environments. However, the implicit assumption of continous (and infinite) nested flux surfaces invoked in deriving such states can be relaxed. The resulting formulation, multi-region relaxed magnetohydrodynamics (MRxMHD), has proven to be successful in many contexts as noted in the Introduction.

Despite the great utility of MRxMHD, especially upon the inclusion of flow, it is still reliant on a variational principle that assumes the invariance of the magnetic and cross helicities, which are ideal MHD invariants. In this study, we have generalized MRxMHD further by adopting the framework of Hall MHD and invoking the magnetic helicity and the \emph{canonical} helicity as the invariants in constructing our variational principle. The presence of the Hall term introduces some mathematical subtleties, given that Hall MHD retains residual two-fluid effects: one of them is manifest in the fact that the canonical vorticity $\nabla \times {\bf P} = {\bf B} + d_i \nabla \times {\bf V}$ is advected along the \emph{ion} trajectory, whilst the magnetic field is advected along the \emph{electron} trajectory, as pointed out in \cite{KLMWW14,DAML16,LMM16}.

After going through the requisite algebra, we arrive at the final results, viz. the partially relaxed states given by (\ref{bVari}), (\ref{VelVari}) and (\ref{rhoVari}), and the interface condition (\ref{interface}). If we consider the incompressible limit of the former trio of equations, and assume that $\Omega_l \rightarrow 0$, the generalizations of the famous double Beltrami states \cite{MY98} are duly obtained. Thus,  MRxHMHD (MRxMHD with Hall effects) plays an analogous role to MRxMHD since the former leads to states akin to the double Beltrami states whilst the latter yields the Woltjer-Taylor (single Beltrami) states. As the double Beltrami states have proven to be fairly successful in both fusion \cite{MY00,YMOIS01,GMY05,Yo10,Aul11} and astrophysics \cite{OSYM01,MMNS01,KM04,LingMa15,LB16,LinB16}, it is natural to suppose that the MRxHMHD equilibria will also prove to be useful in modelling the same phenomena.

We have also analyzed the MRxMHD equilibria obtained in \cite{DHDH14}, which form a subset of the equilibria derived in this paper. We showed that the partially relaxed states with flow that emerge from the Eulerian variational principle are a valid and meaningful subset of ideal MHD equilibria - a fact that lends further credence to our variational principle, and that of \cite{DHDH14}. We also compared these results against the alternative approach espoused in \cite{DYBH15}, which gave rise to a different set of results, and indicated the potential factors that may be responsible for this outcome.  

In subsequent studies, we hope to pursue some promising lines of approach. Our possible avenue is to extend this procedure to encompass electron inertia \cite{AKY15,LMM15,LMM16,DAML16} and/or gyroviscosity \cite{MLA14,LM14}. From the standpoint of applications, we intend to employ the partially relaxed states derived in this paper to study systems where Hall effects play a role; one such example is to extend the approach presented in \cite{SOY05} to study the magnetospheres of the Jovian planets. Secondly, we are in the process of improving the successful SPEC code \cite{Het12} to include flow, which can then be used to study a wide range of issues in fusion plasmas. Lastly, our paper has been centred around an Eulerian approach that yields the static (relaxed) states. It is straightforward to generalize our results to model the dynamical behavior, along the lines of \cite{DYBH15}, by adopting the Lagrangian action principle for Hall MHD that was recently developed in \cite{DAML16}.

\section*{Acknowledgments}
M.L. was supported by NSF Grant No. AGS-1338944 and DOE Grant No. DE-AC02-09CH-11466. H.M.A. would like to thank the Egyptian Ministry of Higher Education for supporting his research activities. H.M.A. wishes to acknowledge the hospitality of M.L., S.R.H., and the Princeton Plasma Physics Laboratory during the course of his visit. The authors thank Profs. Amitava Bhattacharjee, Robert Dewar and Zensho Yoshida for their insightful remarks and encouragement. 


%

\end{document}